\documentclass[a4paper]{jpconf}
\usepackage{graphicx}

\newcommand{\pta}{p_{\mbox{T}}}
\newcommand{\pp}{pp}
\newcommand{\pPb}{p--Pb}
\newcommand{\Pbp}{Pb--p}
\newcommand{\PbPb}{Pb--Pb}

\begin{document}
\title{Recent results on two-particle correlations in ALICE}

\author{Alice Ohlson for the ALICE Collaboration}

\address{CERN, Geneva, Switzerland}

\ead{alice.ohlson@cern.ch}

\begin{abstract}
The discovery of correlations between particles separated by several units of pseudorapidity in high-multiplicity \pp{} and \pPb{} collisions, reminiscent of structures observed in \PbPb{} collisions, was a challenge to traditional ideas about collectivity in heavy ion collisions.  In order to further explore long-range correlations and provide information to theoretical models, correlations between forward trigger muons and mid-rapidity associated hadrons were measured in \pPb{} collisions at $\sqrt{s_{\mbox{NN}}} = 5.02~\mbox{TeV}$.  The results demonstrate that the nearside and awayside ridges extend to $\Delta\eta \sim \pm 5$ and that the $v_2$ of muons, obtained from subtracting the correlation functions in high- and low-multiplicity events, is $(16\pm6)\%$ higher in the Pb-going than in the p-going direction.  The results are compared with AMPT simulations.  
\end{abstract}

\section{Introduction}

Two-particle angular correlations are used to study many aspects of the physics of heavy ion collisions, in particular jet fragmentation and collective effects.  The distributions in relative azimuthal angle ($\Delta\varphi = \varphi_{trig}-\varphi_{assoc}$) and relative pseudorapidity ($\Delta\eta = \eta_{trig}-\eta_{assoc}$) between trigger and associated particles are constructed to obtain the per-trigger yield, $\frac{1}{N_{trig}} \frac{d^2N}{d\Delta\varphi d\Delta\eta}$.  

Correlation functions in collisions such as \pp{} show characteristic features attributed to jet production: a nearside peak localized around $(\Delta\varphi,\Delta\eta) = (0,0)$, representing pairs where the trigger and associated particles are fragments of the same jet, and the awayside peak localized around $\Delta\varphi = \pi$ but extended in $\Delta\eta$, representing pairs in which the trigger and associated particles are in back-to-back jets.   In heavy ion collisions, the same jet features are observed, in addition to structures around $\Delta\varphi = 0$ and $\Delta\varphi = \pi$ extended in $\Delta\eta$.  These `ridges' are often attributed to hydrodynamic flow behavior in the quark-gluon plasma (QGP) and quantified by the Fourier coefficient, $v_2$.  

It was therefore surprising when a nearside ridge was observed in high multiplicity collisions of small systems, \pp{}~\cite{ppridge} and \pPb{}~\cite{pPbridge}.  Subtraction of the correlations in low-multiplicity events from the high-multiplicity correlation functions revealed a symmetric ridge on the awayside~\cite{doubleridge,doubleridgeATLAS}, which is reminiscent of the ridges attributed to flow in heavy ion collisions.  This `double ridge' structure can be quantified by the parameter $v_2$, and further studies demonstrated that the $v_2$ in \pPb{} shows similar mass ordering as was observed in \PbPb{} collisions~\cite{PIDridge,PIDridgeCMS}.  It is important to note that the physical mechanism leading to a non-zero $v_2$ is still under theoretical debate, and the presence of $v_2$ does \emph{not} necessarily imply the existence of hydrodynamics or a QGP in small systems.  

Several theoretical explanations have been proposed to explain the ridge in \pp{} and \pPb{} collisions, such as higher-order glasma graphs within a Color Glass Condensate (CGC) picture, hydrodynamics in small systems, and others (see for example \cite{hydroridge,cgcridge}).  In order to differentiate between the models and constrain future theoretical calculations, it would be important to measure the ridge to larger $\Delta\eta$, as well as to measure the dependence of $v_2$ on the pseudorapidity of the trigger particle.  Both of these points are addressed in the muon-hadron analysis performed in ALICE~\cite{muh}, which will be discussed in detail in the following.

\section{Muon-hadron correlations -- Analysis details}

In the analysis described below, the correlation functions between muons at forward rapidities and charged hadrons at mid-rapidity are constructed in order to investigate the long-range behavior of the double ridge structure for $-5 < \Delta\eta < -1.5$.  Two beam configurations made it possible to measure the $v_2$ coefficients for muons going in the direction of the proton beam and in the direction of the lead ion beam.  

\subsection{ALICE detector}

Proton-lead collisions were recorded with the ALICE detector~\cite{ALICEdet}.  The main subsystems used in the muon-hadron analysis were: the Forward Muon Spectrometer (FMS), the Inner Tracking System (ITS), the Time Projection Chamber (TPC), and the V0 system.  

Trigger muons were detected within the FMS, which has a pseudorapidity coverage of $-4 < \eta_{lab} < -2.5$.  The composition of parent particles of the detected muons depends on transverse momentum ($\pta$): at low $\pta$ the muons predominantly come from weak decays of pions and kaons, while at high $\pta$ the muons are largely the result of heavy flavor decays.  

Two sets of associated hadrons are used.  (1) ``Tracks:'' charged tracks are fully reconstructed in the ITS and TPC, and have a transverse momentum $0.5 < \pta < 4~\mbox{GeV}/c$ and uniform coverage in $0 < \varphi < 2\pi$ and $|\eta| < 1$.  (2) ``Tracklets:'' short track segments are reconstructed in the Silicon Pixel Detector (SPD), the two innermost layers of the ITS.  The tracklet momentum is correlated with the differences of the azimuthal ($\Delta\varphi_h$) and polar angles ($\Delta\theta_h$) of hits in the SPD layers, and so a tracklet sample with $\langle\pta\rangle \approx 0.75~\mbox{GeV}/c$ is obtained by requiring that $\Delta\varphi_h < 5~\mbox{mrad}$.  

The multiplicity of charged particles incident on the V0 detectors is used to classify the overall event activity.  Only two of the four rings in each V0 detector are used in order to achieve symmetric coverage: the innermost two rings of the V0C ($-3.7 < \eta_{lab} < -2.7$) and the outermost two rings of the V0A ($2.8 < \eta_{lab} < 3.9$).  

\subsection{Collected data sets}

During 2013 the LHC provided proton and lead beams at beam energies of $4~\mbox{TeV}$ and $1.58~\mbox{ATeV}$, respectively, for a center-of-mass collision energy of $\sqrt{s_{\mbox{NN}}} = 5.02~\mbox{TeV}$.  Because of the unequal beam energies, the center-of-mass rapidity is shifted by 0.465 units in the direction of the proton beam (in the following all pseudorapidities will be quoted in the laboratory frame, $\eta_{lab}$).  The beams were provided in both configurations: \pPb{}, with the proton beam going in the direction of the muon arm, and \Pbp{}, with the lead beam facing into the muon arm.  

Muon-tracklet correlations were investigated in both \pPb{} and \Pbp{} collisions.  Because of the triggering scheme chosen during the \Pbp{} data-taking period, the TPC was not read out in the trigger cluster for the majority of the data set, and therefore muon-track correlations were studied only in \pPb{} data.  

\subsection{Analysis methodology}

\begin{figure}[t]
\centering
\begin{minipage}{.25\textwidth}
  \centering
  \includegraphics[width=.95\linewidth]{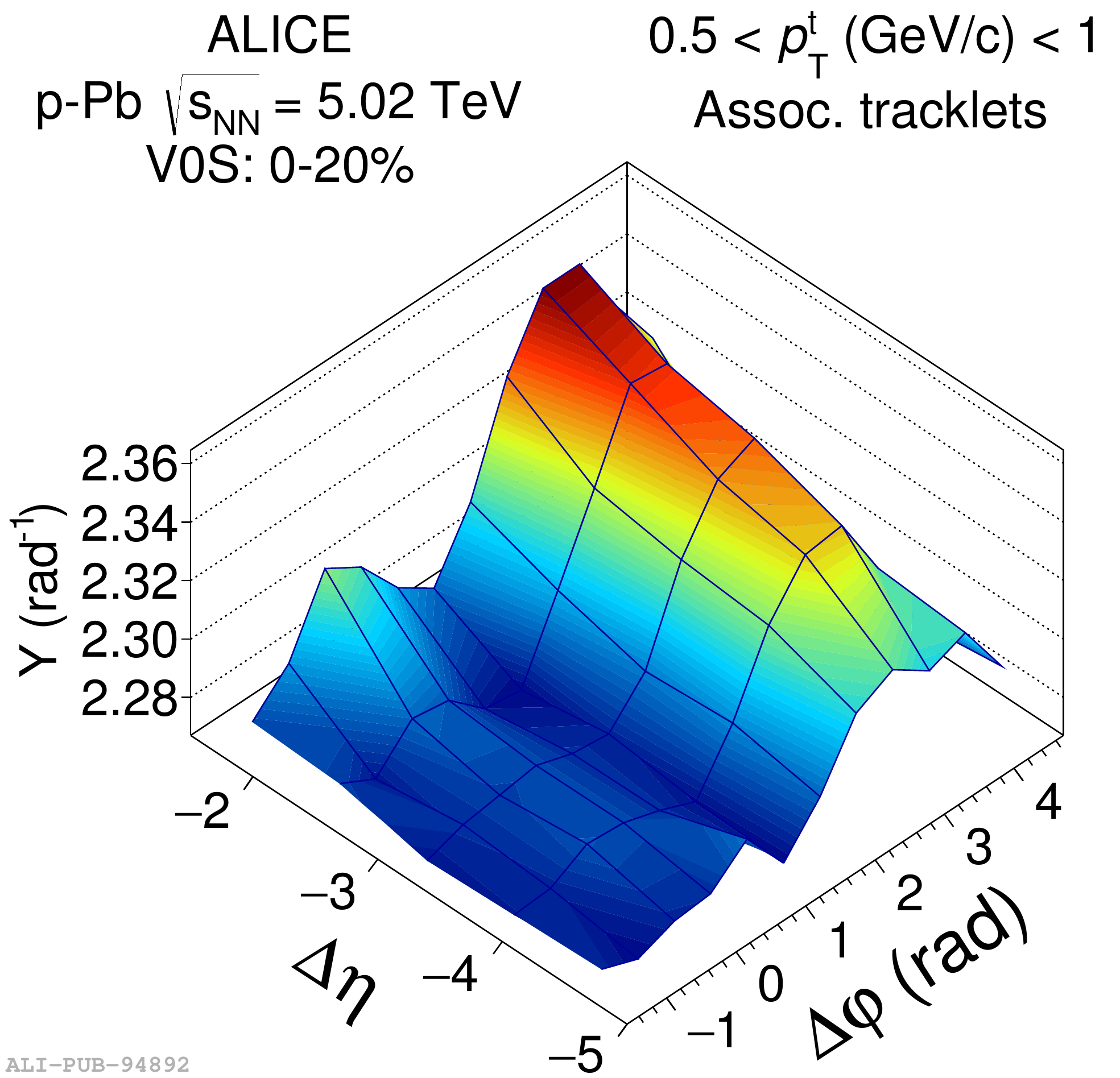}
\end{minipage}%
\begin{minipage}{.25\textwidth}
  \centering
  \includegraphics[width=.95\linewidth]{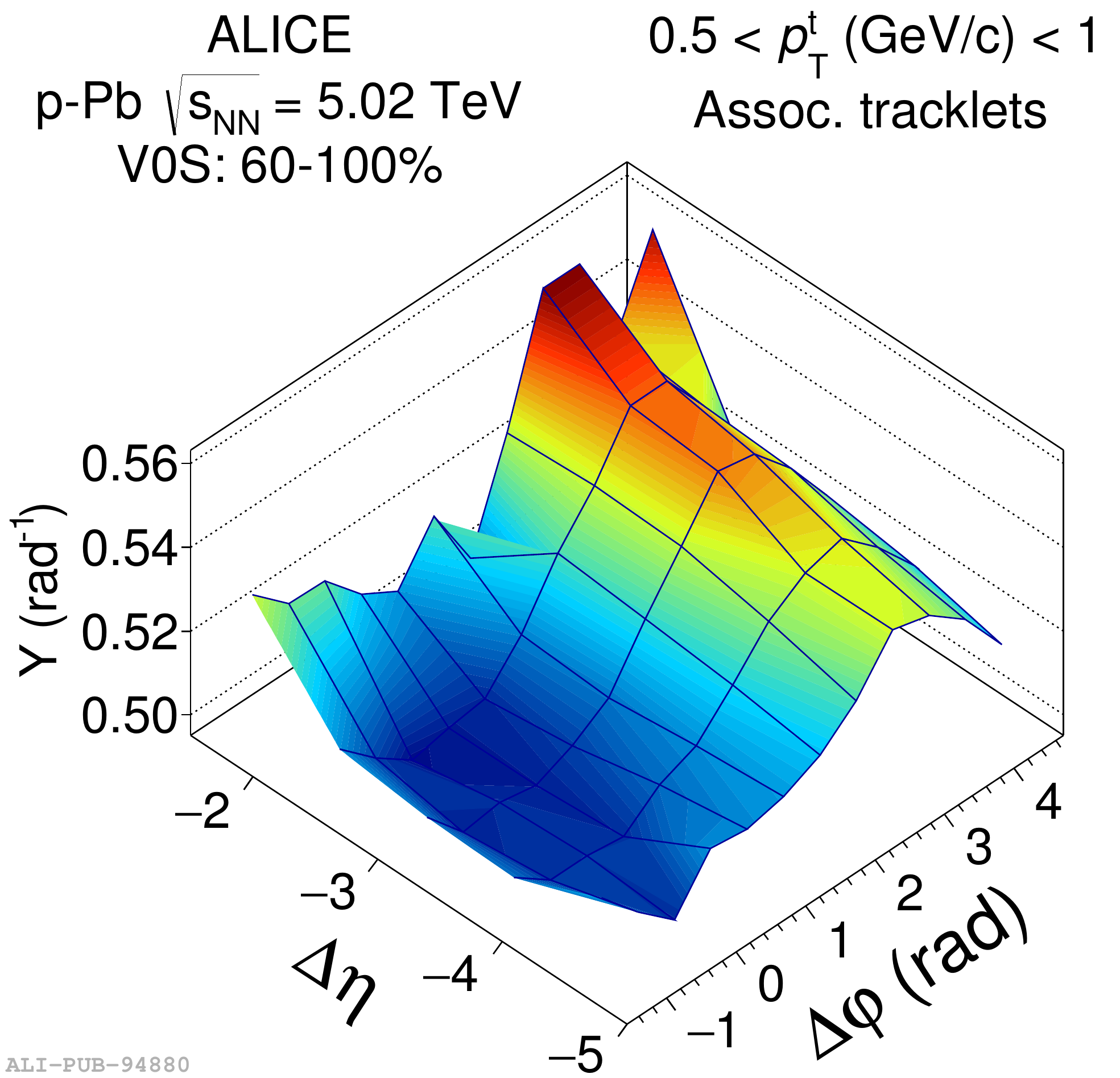}
\end{minipage}%
\begin{minipage}{.25\textwidth}
  \centering
  \includegraphics[width=.95\linewidth]{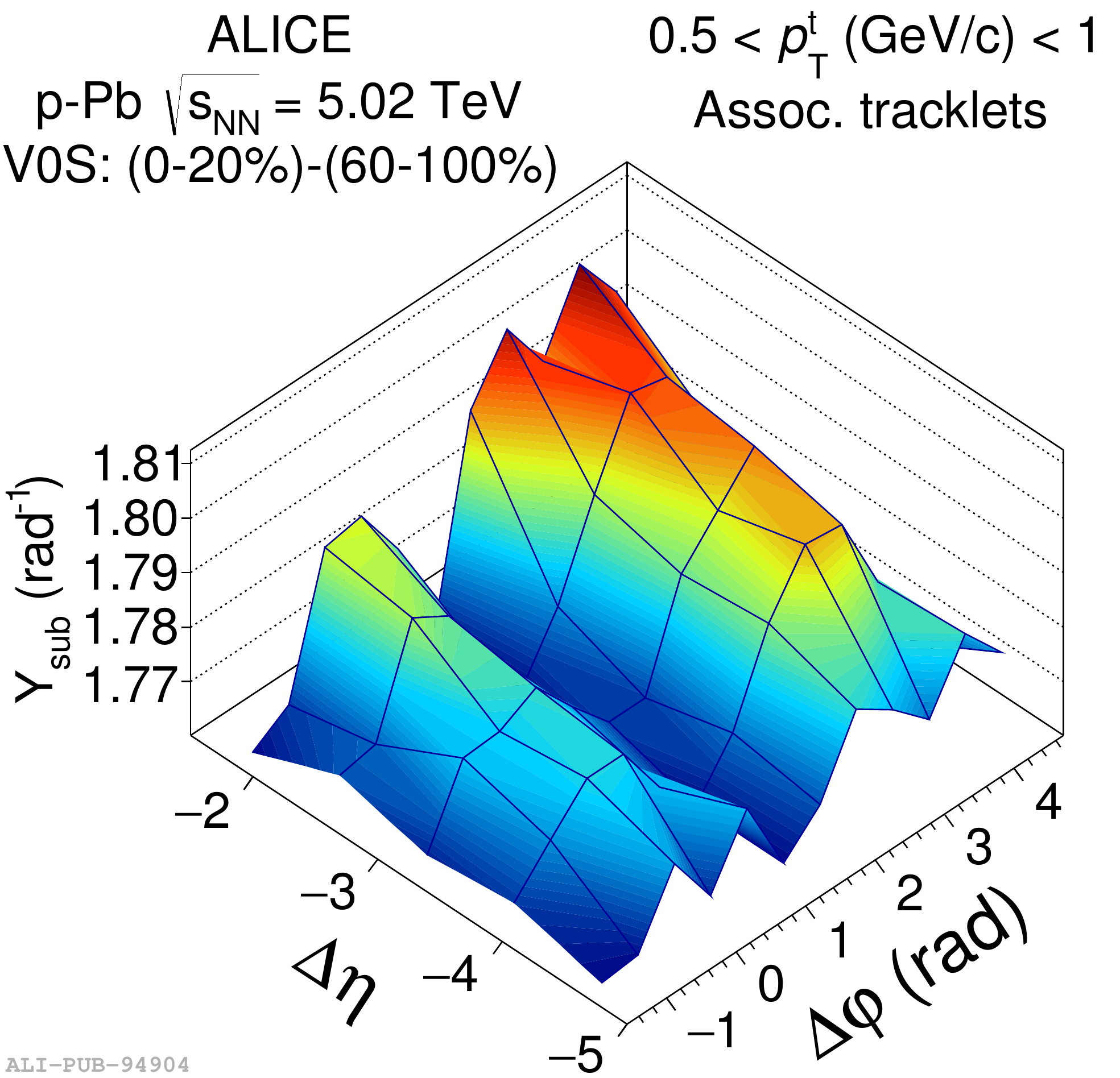}
\end{minipage}%
\begin{minipage}{.25\textwidth}
  \centering
  \includegraphics[width=.95\linewidth]{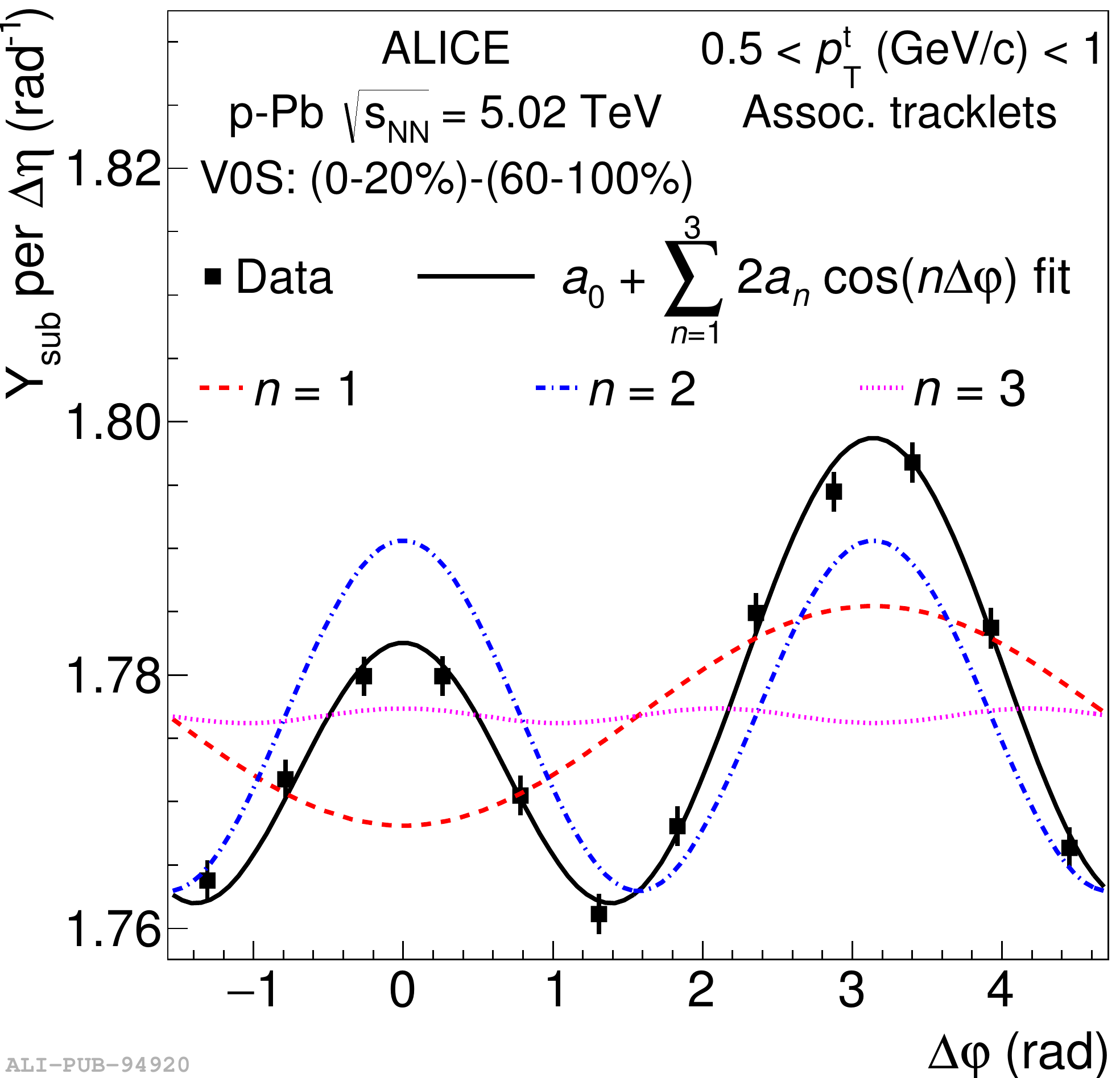}
\end{minipage}%
\caption{Muon-tracklet correlations in \pPb{} collisions.  From left to right: the correlation function in high-multiplicity events in which the nearside ridge is clearly visible, the correlation function in low-multiplicity events where there is no obvious ridge, the subtracted correlations which show the double ridge structure, and the $\Delta\varphi$ projection with a Fourier cosine series fit.}
\label{fig:subt}
\end{figure}

The muon-track and muon-tracklet correlation functions were measured in high-multiplicity (the top 20\% of the analyzed event sample) and in low-multiplicity (60-100\%) events.  As in~\cite{doubleridge}, the low-multiplicity correlations are subtracted from the high-multiplicity correlation functions to remove structures associated with jet fragmentation.  After subtraction, the correlation functions are projected onto $\Delta\varphi$, and then fit with a Fourier cosine series, 
\begin{equation}
a_0+\displaystyle\sum_{n=1}^{3} 2a_n\cos(n\Delta\varphi).
\end{equation}
From the coefficients of the Fourier series, the pair azimuthal anisotropy $V_{2\Delta}\{\mbox{2PC,sub}\}$ can be obtained:
\begin{equation}
V_{2\Delta} \{\mbox{2PC,sub}\} = \frac{a_2}{a_0+b},
\end{equation}
where $b$ is the baseline in the low-multiplicity event class.  Assuming factorization of the $v_2$ values for trigger and associated particles, the $v_2$ of trigger muons can be obtained by dividing $V_{2\Delta} \{\mbox{2PC,sub}\}$ by the $v_2$ of associated tracks or tracklets, which is obtained from track-track and tracklet-tracklet correlations, respectively:
\begin{equation}
v_2^{\mu}\{\mbox{2PC,sub}\} = \frac{V_{2\Delta}^{\mu-h} \{\mbox{2PC,sub}\}}{v_2^h} = \frac{V_{2\Delta}^{\mu-h} \{\mbox{2PC,sub}\}}{\sqrt{V_{2\Delta}^{h-h} \{\mbox{2PC,sub}\}}}.
\end{equation}
The agreement between $v_2^{\mu}\{\mbox{2PC,sub}\}$ measured in muon-track and muon-tracklet correlations (which have very different $\pta$-distributions for the associated hadrons) demonstrates that the factorization assumption is valid.  An example set of correlation functions from the muon-tracklet correlations analysis in \pPb{} collisions is shown in Fig.~\ref{fig:subt}.  

\section{Muon-hadron correlations -- Results and model comparisons}

\begin{figure}
\centering
\begin{minipage}{.46\textwidth}
  \centering
  \includegraphics[width=.95\linewidth]{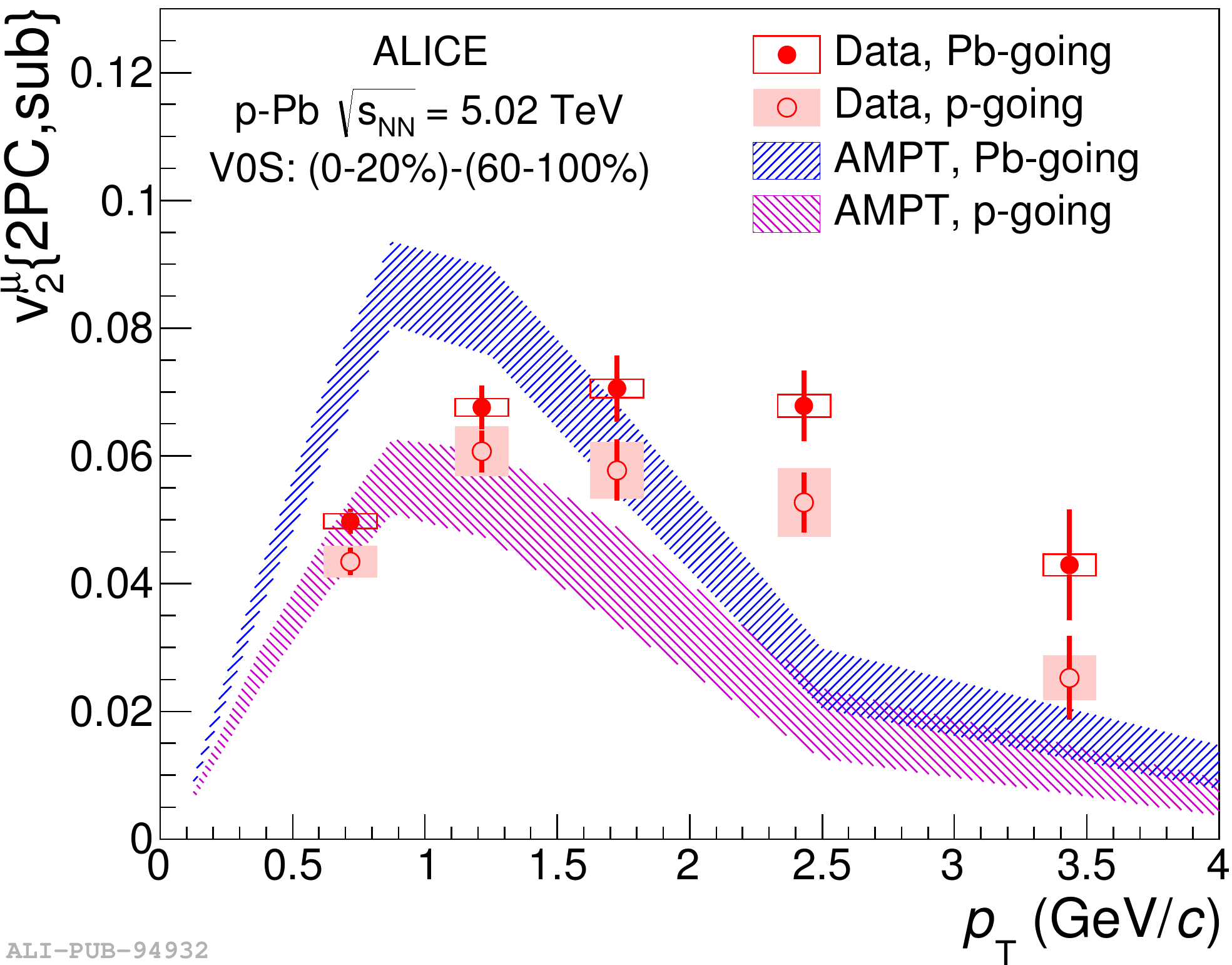}
  \caption{$v_2^{\mu}\{\mbox{2PC,sub}\}$ in the Pb-going and p-going directions as a function of $\pta$ compared to AMPT.}
  \label{fig:v2mu}
\end{minipage}%
\hfill
\begin{minipage}{.46\textwidth}
  \centering
  \includegraphics[width=.95\linewidth]{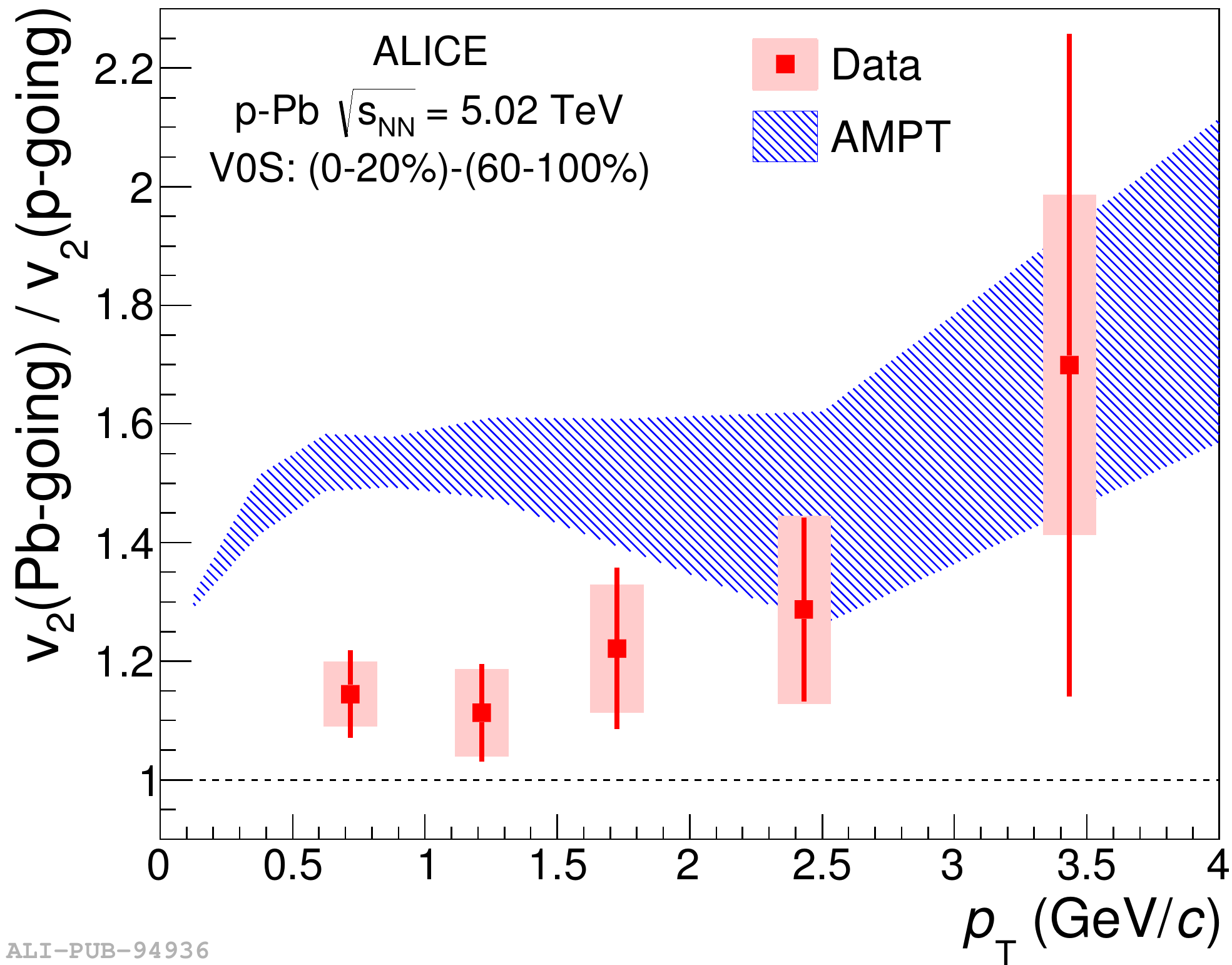}
  \caption{Ratio of $v_2^{\mu}\{\mbox{2PC,sub}\}$ in the Pb-going and p-going directions as a function of $\pta$ compared to AMPT.}
  \label{fig:v2ratio}
\end{minipage}
\end{figure}

The $v_2^{\mu}\{\mbox{2PC,sub}\}$ of muons has been measured as a function of $\pta$ in the p- and Pb-going directions, and the results are shown in Figs.~\ref{fig:v2mu} and~\ref{fig:v2ratio}.  It is observed that $v_2^{\mu}\{\mbox{2PC,sub}\}$ in the Pb-going direction is consistently higher than $v_2^{\mu}\{\mbox{2PC,sub}\}$ in the p-going direction.  The ratio, shown in Fig.~\ref{fig:v2ratio}, is roughly independent of $\pta$ within statistical and systematic uncertainties.  A constant fit to the ratio yields $1.16\pm0.06$.  

The data are compared with an AMPT~\cite{refAMPT} simulation in which the muon decay products are scaled to account for the efficiency of the absorber in the ALICE FMS.  In Fig.~\ref{fig:v2mu} it is seen that while AMPT qualitatively describes the $\pta$-dependence at low $\pta$, there are significant quantitative differences in the $\pta$-dependence and $\eta$-dependence between data and the model.  At high $\pta$ (above $\pta \sim 2~\mbox{GeV}/c$), where muon production is dominated by heavy flavor decays, AMPT does not describe the data well.  This could be because heavy flavor muons have a non-zero $v_2$, or the parent particle composition or $v_2$ values in data and AMPT are different.  

Theoretical treatments of forward-central muon-hadron correlations indicate that the observables measured here are sensitive to the dynamics and possible collective effects in small systems~\cite{theorycomp}.  However, current theoretical calculations cannot be directly compared with experimental results, because the effects of the absorber are included in the experimental data (unfolding such effects could not be done in a model-independent way).  Future model calculations should use the efficiencies provided in~\cite{muh} in order to compare directly to the experimental results.  

\section{Conclusions}
Correlations between forward trigger muons and mid-rapidity associated hadrons can provide insight into the $\eta$-dependence of long-range correlations in \pPb{} collisions.  The analysis is performed using muons detected in the FMS and charged hadrons reconstructed in the central barrel in ALICE in \pPb{} and \Pbp{} collisions at $\sqrt{s_{\mbox{NN}}} = 5.02~\mbox{TeV}$.  The results demonstrate that the nearside and awayside ridges extend to $\Delta\eta \sim \pm 5$ with trigger particles out to $|\eta| \sim 4$.  Furthermore, the $v_2^{\mu}\{\mbox{2PC,sub}\}$ of muons is $(16\pm6)\%$ higher in the Pb-going direction than in the p-going direction, and this ratio is roughly independent of momentum.  While there is qualitative agreement between data and AMPT at low $\pta$, the model does not yet quantitatively describe the data.  Future measurements and model comparisons will allow us to gain understanding of the production mechanisms of long-range correlations in small systems.

\section{References}
\bibliographystyle{iopart-num} \bibliography{sqm_proc_bib}

\end{document}